\documentclass[superscriptaddress]{revtex4-1}
\usepackage{hyperref}

\usepackage{graphicx}
\usepackage{natbib}

\newcommand\bb{\begin{eqnarray}}
\newcommand\ee{\end{eqnarray}}

\newcommand\bs{\boldsymbol}

\newsavebox{\astrutbox}
\sbox{\astrutbox}{\rule[-5pt]{0pt}{20pt}}

\newcommand\etal{\mbox{{\it et al.}\;}}

\usepackage{color}

\newcommand\et{{\it et}\ }
\newcommand\tmu{\tilde{\mu}}
\newcommand\tC{\tilde{C}}
\newcommand\bmnabla{\bs{\nabla}}
\newcommand\vectr{\mathbf{r}}

\usepackage{amsmath}
\usepackage{amssymb}

\begin{document}

\title{Transition to chaos in wave memory dynamics in a harmonic well :\\ deterministic and noise-driven behaviour}
\date{?; revised ?; accepted ?. - To be entered by editorial office}

\author{S. Perrard}
\affiliation{Laboratoire FAST, CNRS UMR 7608, CNRS, Universit\'e Paris-Saclay, 91405 Orsay, France, EU}
\affiliation{LadHyX, CNRS UMR 7646, École Polytechnique, 91128, Palaiseau, France, EU}
\affiliation{Current affiliation: D\'epartement de Physique ENS, PSL Universit\'e, CNRS, 24 rue Lhomond, 75005 Paris.}
\author{M. Labousse}
\affiliation{Gulliver, CNRS UMR 7083, ESPCI Paris and PSL University, 75005 Paris France, UE}


\begin{abstract}
A walker is the association of a sub-millimetric bouncing drop moving along with a co-evolving Faraday wave. When confined in a harmonic potential, its stable trajectories are periodic and quantised both in extension and mean angular momentum. In this article we present the rest of the story, specifically the chaotic paths. They are chaotic and show intermittent behaviours between unstable quantised set of attractors. First we present the two possible situations we find experimentally. Then we emphasise theoretically two mechanisms that lead to unstable situations. It corresponds either to noise-driven chaos or low-dimensional deterministic chaos. Finally we characterise experimentally each of these distinct situations. This article aims at presenting a comprehensive investigation of the unstable paths in order to complete the picture of walkers in a two dimensional harmonic potential.
\end{abstract}

\maketitle

{\bf A {\it walker} is the association of a bouncing drop and its associated guiding Faraday wave field. This dual system is one of the rare experimental situation encoding a pilot wave dynamics. It has shown multiple behaviours driven by temporal non local effects and reminiscent of one particle quantum-like behaviours. When confined by a harmonic potential, its stable solutions form a set of attractors that are quantised both in extension and mean angular momentum. This striking results has been observed experimentally and numerically by different groups. Here we investigate the case of unstable solutions presenting intermit behaviours between the footprint of the aforementioned set of attractors. We present here a detailed and unified picture of these chaotic trajectories. We show that at least two distinct mechanisms arise and investigate their theoretical origin. The first mechanism is the signature of a low-dimensional chaos as previously reported by several groups. The second mechanism is driven by a multiplicative noise encoded within the multiple wave degrees of freedom. Finally we characterise experimentally these two mechanisms and present a unified and synthetic picture of all intermittent transitions that have been documented so far.}


\section{Introduction}

	The transition from deterministic dynamics toward highly disordered, unpredictable dynamics has been thoroughly studied in physics. The dynamics of a walker~\cite{Walker_Nature}, the association of a bouncing drop with the wave field it generates, also shows a transition from regular, periodic trajectories toward disordered states under confinement. The specificity of walker dynamics is that it exhibits wave-like behaviours and quantised sets of attractors both in the deterministic regime~\cite{Fort_PNAS,Perrard_Nature_2014,JBAnnRev} or in the highly disordered regime~\cite{Harris_PRE_2013,Harris_JFM_2014}. The study of the transient from one regime to another has been less studied, but it has now been reported experimentally~\cite{Perrard_PRL_2014} and numerically~\cite{Tambasco_2016}. Other studies have also reported the arising of more complex trajectories~\cite{ExoticOrb} that are located in this intermediate regime. The origin of the instability mechanisms driving the system from stable to disorder are still unclear. However the use of standard dynamical system tools (bifurcation diagram, first return map, period doubling research) have drawn a transition to low-dimensional chaos which implies a loss of determinism by increasing lack of predictability. Tambasco~\textit{et al.}~\cite{Tambasco_2016} in particular have compared three confinement configurations (Coriolis force, central force and Coulomb potential) to show that several scenarii of transition to chaos can be observed, depending on the confinement type. \\
	
	In the present article, we go beyond our first report of low dimensional chaos~\cite{Perrard_PRL_2014} both experimentally and theoretically. Experimentally, we base our analysis on data obtained with a central force field. We provide evidence of a transition to chaos from periodic orbits to unpredictable dynamics of two types. One type follows the previous observation of Perrard~\textit{et al.}~\cite{Perrard_PRL_2014} and the previous detailed analysis of Tambasco~\textit{et al.}~\cite{Tambasco_2016}. The transition is in particular now thoroughly characterised and consistent between authors for small extension orbits. In contrast, for larger orbit size, the experiments show a total loss of determinism while the transition between reminiscent periodic trajectories are still deterministic. It is in favour of a noise amplification mechanism throughout information storage in the wave field. This article is organised as follows. In Sec.~~\ref{expchaos}, we present the experimental setup and summarise the main results. In Sec.~\ref{phasecontract}, we derive theoretically a set of equation in the continuous limit, that enables us to compute the global Lyapunov exponent of the walker dynamics. The result is not specific to the central force field: we show that any volume of phase space globally contracts at all memory but it converges toward zero in the infinite memory limit. This constrain roots a route towards noise-driven chaos. Then we provide in Sec.~\ref{geoconstrain} an analysis of periodic orbits and show that a constraint on the mode amplitudes applies. It evidences the number of degree of freedom that are relevant to describe the dynamics, depending on the spatial extension of walker trajectory. Finally we characterise experimentally these two scenarios in Sec.~\ref{expchar}. For the sake of simplicity, we start with low-dimensional chaotic type in Sec.~\ref{chaosdet} and emphasise the noise amplification mechanism in Sec.~\ref{noisechaos}.

\section{Experimental set up and evidence of chaos~\label{expchaos}}
\subsection{Methods\label{methods}}

The experimental setup consists in a bath filled with silicon oil of viscosity $\nu = 20$~cp shaken vertically at an acceleration $\gamma = \gamma_0 \cos(2 \pi f t)$ at a frequency $f$= 80~Hz. The acceleration amplitude $\gamma_0$ of the bath is tuned in the vicinity below the Faraday acceleration threshold $\gamma_F\simeq 4.5$ g above which standing Faraday waves are spontaneously generated~\cite{FaradayOriginal,Douady_JFM,JFMKumarTuckerman}. The control parameter for the damping time of the drop generated waves is $M = \gamma_0 / (\gamma_F - \gamma_0)$ which ranges typically from 10 to 200. A sub-millimetric drop of mass $m_D$ bounces on the bath on a doubling period regimes at the Faraday period $T_F$, as described in~\cite{JFM_Suzie}. In this situation the drop is self-propelled by a dynamically co-moving localised standing waves of wavelength $\lambda_F=4.5$~mm~\cite{Walker_Nature}.  The dynamics of the drop for increasing memory is explored in a confined situations, in which the drop is submitted to an external force. This force is generated by the use of magnetic fields. Two coils placed in a Helmholtz configuration around the liquid bath generate a homogeneous vertical magnetic field $B_0$ in the plane of the unperturbed liquid surface. Magnetic field gradients are generated with a cylindrical magnet made of an Neodyme Bore alloy placed at a tunable distance $d$ of the liquid surface. The total magnetic field is therefore $\bs B = B_0 \bs e_z + \bs B_m$ where ${\bs B_m}$ is the magnetic field generated by the magnet. The drop is made magnetic sensitive by encapsulating inside a droplet of ferrofluid of mass $\approx$0.05~$m_D$. The potential energy of interaction writes $E_m \propto - \bf m . \bf B$ where $\bf m \propto B$ is the induced magnetic dipole. Our ferrofluid was previously characterised in~\cite{Browaeys_thesis} and the entire magnetic set up calibrated using a magnet oscillating horizontally. The full experimental details are given in~\cite{Perrard_Nature_2014}. In the central region of interest ($r < 3 \lambda_F$), we are left with a harmonic potential for the drop where the magnetic force per unit mass $\bf{F}$ writes :
\begin{equation}
{\bf F}= - \omega^2 \vectr
\end{equation}
Here $\omega = \sqrt{k/m_W}$ is the magnetic frequency and $\vectr$ is the radial vector whose length is the distance to the center of the system. A correction to the drop inertia originating from the wave field has been identified experimentally in the central force set up as well as on a rotating bath~\cite{Fort_PNAS,Oza_JFM_2_2013} and theoretically rationalised~\cite{Boosht}. The deviation from $m_W=m_D$ was then measured throughout the radius $R$ of the stable circular motion of the confined drop. Eventually, we calibrate the system directly on circular orbits. In the following we will use the non dimensional confinement parameter $\Lambda = V_0/ (\omega \lambda_F)$  that sets the orbit size where $V_0$ is the mean speed of the walker. For circular motions and no significant additional force generated by the wave field we observe $R/\lambda_F = \Lambda$. The dynamics is further explored as a function of two main parameters. The memory parameter $M$ (or memory time $\tau=M T_F$) sets the temporal coherence of the system. It is tuned throughout the vertical acceleration of the path. The confinement parameter $\Lambda$ sets indirectly the mean extension of the trajectories. $\Lambda$ is varied by changing the distance $d$ from the magnet to the bath. We then explore different type trajectories of increasing complexity by increasing $\Lambda$.
	
\begin{figure}
\centering
\includegraphics[width=0.8\columnwidth]{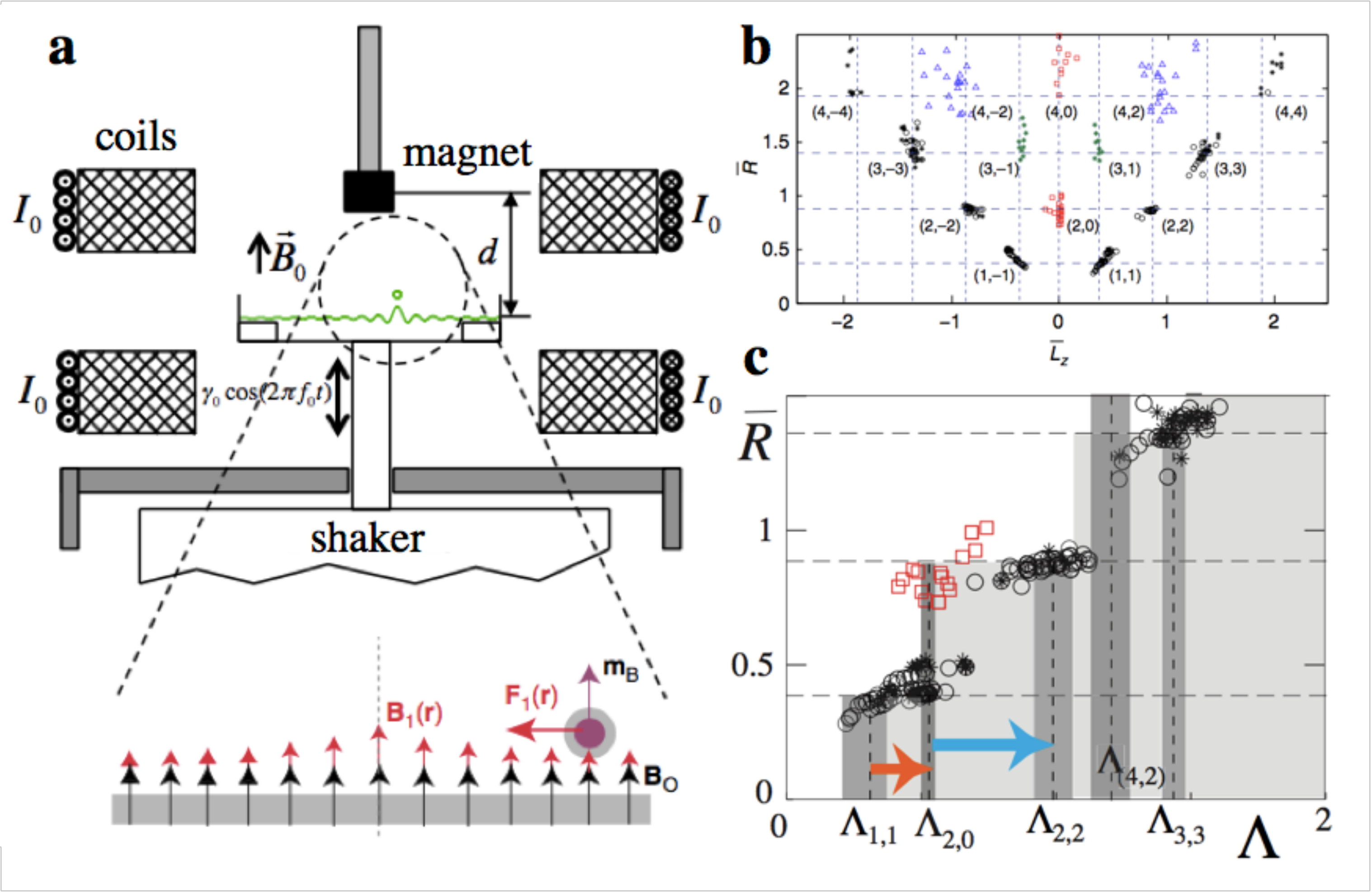}
\caption{{\bf Experimental set up and region of interest}. a) Sketch of the experimental set up. A drop of silicon oil encapsulating ferrofluid (zoomed in view) is generated on a vertically vibrating liquid bath.  Immersed in a magnetic field $B$ generated by Helmholtz coils ($B_0$ field) and a cylindrical magnet ($B_m$ field) the drop is submitted to a central attractive harmonic force. The drop exhibit  b) Diagram of spatial extension $\bar R$ and angular momentum $\bar L$ computed on one period of motion for all observed periodic trajectories from~\cite{Perrard_Nature_2014}. c) Zoom in view on the mean spatial extension $\bar R$ of trajectories as a function of the confinement $\Lambda$ for $50<M<100$. Regions of stable periodic orbits (dark grey) and regions of intermittent behaviour (light grey). We focus here on the light grey regions. Figure~1b adapted from~\cite{Perrard_Nature_2014} with courtesy of Perrard~\textit{et al.}}
\label{fig1}
\end{figure}

\subsection{Stable \& unstable periodic orbits in a harmonic potential}

The stable trajectories generated by a walker in a harmonic well have been reported experimentally using the set up previously described~\cite{Perrard_Nature_2014,Perrard_PRL_2014}, numerically using discrete time models~\cite{Labousse_NJP_2014,Durey_2017} or continuous time model~\cite{Tambasco_2016,Kurianski_2017} and investigated theoretically~\cite{Labousse_NJP_2014,Labousse_PRE_2016,Kurianski_2017,Durey_2017}. The observation of stable trajectories of different symmetries is a common feature, robust to changes of the model details. A quantisation of mean radius $\bar R$ and mean angular momentum $\bar L$ for stable trajectories is also observed by all authors, although differences may arise between the models in particular for the exact quantisation selection rule~\cite{Kurianski_2017} or the accuracy of the angular momentum quantisation~\cite{Durey_2017}. We introduce here the reader to walker's dynamics in a harmonic well throughout the experimental data first reported in~\cite{Perrard_Nature_2014}. The mean radius of the stable orbits observed experimentally in the range $50<M<100$ are represented in figure~\ref{fig1}a as a function of the confinement parameter $\Lambda$. The colour codes the symmetry of the various trajectories: black for circular symmetry, red for two fold symmetry (lemniscate) and blue for 3-fold symmetry (trifolium). All the stable solutions that we found experimentally are classified in a diagram $(\bar{R},\bar{L}_z)$ (figure~\ref{fig1}b) and each cluster of states is labelled accordingly by a set of two integers $(n,m)$. However, these stable orbits are only one side of the full story. As the memory parameter $M$ is increased, most parameter values $\Lambda$ lead to unstable orbits with intermittent shift between trajectories of different symmetry. This feature has been reported experimentally~\cite{Perrard_PRL_2014} and thoroughly studied numerically by Tambasco~\textit{et al.}~\cite{Tambasco_2016}. We propose here to describe the transition from stable to unstable orbit in a more systematic manner than it was previously reported. We will highlight in particular the two types of intermittency that have been experimentally observed.
	
\subsection{Intermittent regimes :definition \& first observation}

	The first destabilisation of a non periodic orbit emerges at intermediate memory ($M \approx 50$) for highly confined trajectories ($\Lambda \approx 0.5$). This regime corresponds to a dimensionless memory length $S_{\textrm{Me}}=V_0 \tau /\lambda_F$ of order unity. Figure~\ref{fig1}c shows a close up of mean radius trajectory as a function of the confinement parameter $\Lambda$ for the modes $n=1$ and $n=2$ for $\textrm{M}=50$. The stable zones corresponding to simple periodic orbits have been highlighted in dark grey. We observe that between two stable regions, no stable orbits are observed. We will here detail the analysis for the observation of the first two regions of non periodic orbits. For this purpose, we define the critical values $\Lambda^{\pm}_{(n,m)}$ of the confinement parameter $\Lambda$ that delimits the stability range of each mode $(n,m)$. Each mode characterised by an integer couple $(n,m)$ is therefore observed in the range $\Lambda^-_{(n,m)} < \Lambda < \Lambda^+_{(n,m)}$. The first zone of interest is located in $\Lambda$ values between circular orbits $n=1$ and lemniscate mode $(n,m) = (2,0)$, hence $\Lambda_{1,1}^+ < \Lambda < \Lambda_{2,0}^-$. The second zone is located between lemniscate mode $(n,m) = (2,0)$ and circular orbits $n=2$, hence $\Lambda_{2,0}^+ < \Lambda < \Lambda_{2,2}^-$. The width of unstable zones increases with $M$, such that the limit of stability $\Lambda^{\pm}_{n,m}$ is a function of memory.  \\
	
\begin{figure}
\centering
\includegraphics[width=0.7\columnwidth]{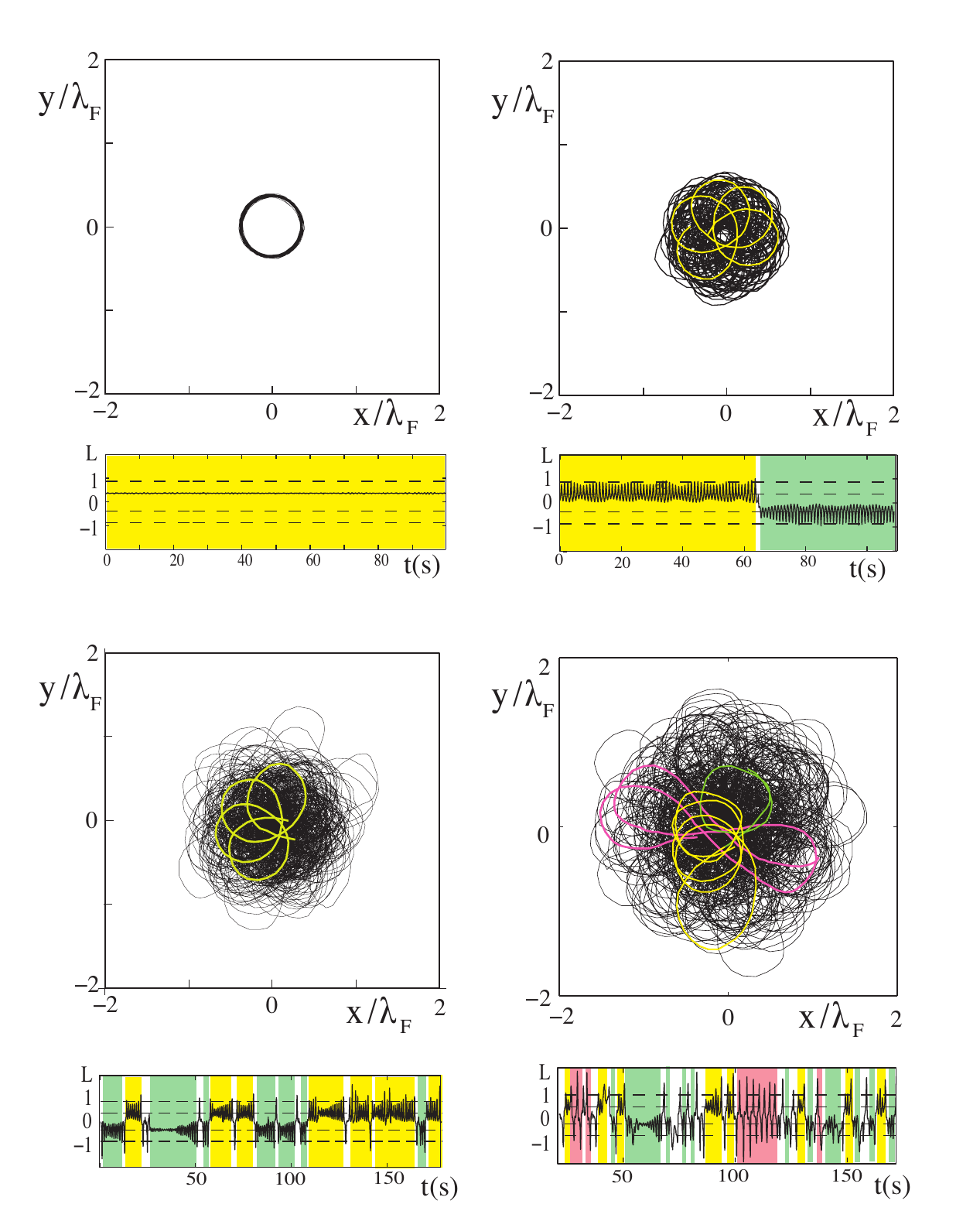}
\caption{{\bf First sign of instability} Experimental trajectories observed for $M=63$ and for increasing values of $\Lambda$ starting from $\Lambda = 0.46$. a) $\Lambda=0.46$. b) $\Lambda=0.49$. c) $\Lambda = 0.5$. d) $\Lambda = 0.62$. Drop velocity $V = 10 \pm 0.5$~mm/s. Bottom: temporal record of dimensionless angular momentum $L$ with color codes for $(n,m)=(1,1)$ (yellow), $(n,m)=(1,-1)$ (green) and $(n,m)=(2,0)$ (green).}
\label{fig2}
\end{figure}
Four experimental trajectories obtained for increasing values of $\Lambda$ are represented in figure~\ref{fig2} for $\textrm{M} = 50$. As the confinement parameter $\Lambda$ increases the circular orbits become unstable. This phenomenon is better observed from a temporal record of the angular momentum $L_z$, as illustrated in figure~\ref{fig2}. It shows a first reversal of rotation direction after a long duration of periodical motion (typically 50 orbital periods $T_o$). For higher values of $\Lambda$, the average time spent in each direction of rotation decreases. The trajectory is then composed of gently diverging oscillations of angular momentum (henceforth \textit{laminar phase}) and abrupt transitions in between (henceforth \textit{chaotic phase}). In the laminar phase, the angular momentum $\bar L_z$ averaged on the period of the orbit is still equal to the unperturbed mode. Time intervals of mean angular momentum $\bar L_z = L_z^{(1)}$ has been coloured in yellow, while $\bar L_z = L_z^{(-1)}$ zones were coloured in green. For higher values of $\Lambda$, a third type of motion emerge corresponding to a large oscillation of $L$. The direct trajectory observation reveals that the transition is mediated by a third building block of motion corresponding to the lemniscate mode (red zone of fig~\ref{fig2}). Yellow, green and red zones of variable length are then sequentially observed. The total time spend in this three types of motion covers most of the recording. The interval spent in one of these modes is called \textit{laminar phase} while the short, erratic motion between these phases is called \textit{chaotic burst}. It is already remarkable that a succession of several eigenmodes of different symmetries appears in an apparent erratic dynamics. \\
	
	The arising of chaos from a stable limit cycle that destabilises spontaneously can be observed in numerous nonlinear dynamical systems as soon as the number of degree of freedom is equal or greater than 3 for continuous non linear systems~\cite{Strogatz_1994}. In this route to chaos several scenario can be identified each of them being associated to different properties of the associated intermittency. They can be sorted out in particular by looking at the distribution of time spend in laminar phases. These routes are described in details in~\cite{Berge_1984}. For the dynamics of a walker, the temporal records of angular momentum show similarity with intermittency of type II and III, in which duration of the laminar phases are unbound. For walkers dynamics it has been observed and characterised numerically by Tombasco et al.~\cite{Tambasco_2016}. The appearance of non commensurate frequency may be associated to type III intermittency which is in practice rarely observed experimentally~\cite{Manneville_2006}. \\
	
	\section{Theoretical origin of wave self-organisation and  different route towards wave chaos. ~\label{theory}}
	
In this Section, we emphasize the theoretical mechanisms leading to the two distinct chaotic transitions. A numerical investigation has been performed by Tombasco et al.~\cite{Tambasco_2016} and indicates also these different routes towards chaos. We present here a complementary point of view by pointing out general mechanisms supporting these different classes of instabilities. We show that these mechanisms arise from fundamental reasons intrinsically due to the wave nature of the system. The first mechanism described in Sec.~\ref{phasecontract} is due the way a wave stores information within an infinite number of degrees of freedom. The second mechanism described in Sec.~\ref{geoconstrain} is a direct consequence of the mathematical constraints arising from periodic orbits. 
 
	\subsection{Phase space contraction and its relation to the memory parameter: a route towards a noise-driven chaotic scenario \label{phasecontract}}
Instabilities and chaotic transitions are usually characterised by analysing the rate of divergence in each direction of phase space or the Lyapunov spectrum $\left\lbrace \lambda_i\right\rbrace _{i \in \mathbb{N}}$. In general this is a difficult task for integro-differential systems as is the walker dynamics where the definition of a phase space can be by itself a tedious task. First, we reformulate the equations of motion into a normal form and we elucidate the mathematical structure of the phase space. Then we show that it is possible to calculate the divergence rate of a small volume of phase space; We relate it to the memory parameter of the system.  Finally we calculate the global Lyapunov $\lambda$. We emphasize important consequences on  system stability originating from the discreteness of the system.
 
\subsubsection{Normal form of the path memory dynamics \label{Normalform}}
We first recall the equations of motion and then reformulate them into a local normal form $\dot{\bm{\mathcal{Z}}}= \bm{\mathcal{F}}\left( \bm{\mathcal{Z}}\right)$ where $\bm{\mathcal{Z}}$ is the state vector of the dynamics and $\bm{\mathcal{F}}$ a local evolution function to specify. The dynamics is driven by a pilot-wave originating from the modulation of the interface. Its footprint is a standing surface $h$ field co-evolving with the drop horizontal motion~\cite{Fort_PNAS}    
\begin{equation}
h=h_0\sum\limits_{k=-\infty}^N J_0\left(k_F\Vert \vectr_N-\vectr_k \Vert \right)e^{-(t_N-t_k)/\tau}, 
\label{Fortwave}
\end{equation}
where $h_0$  the wave amplitude, $k_F=2\pi/\lambda_F$ is the Faraday wave vector, $\vectr_N$ is the drop position at the $N$th impact at time $t_N$. The indexes $k<N$ indicate past situations and $J_0$ is the 0 order Bessel function of first kind. The concept of path memory dynamics has been introduced in its discrete form by Eddi~\textit{et al.}~\cite{Eddi_JFM_2011}. It has been investigated further in its discrete form by Milewski~\textit{et al.}~\cite{Milewski_JFM_2014} and in a continuous form by Oza~\textit{et al.}~\cite{Oza_JFM_1_2013} we shall discuss later. The path memory is an essential feature at the origin of the temporal non-local properties of walker systems. An iterative discrete model based on this path memory has been benchmarked and so far has been able to reproduce quantitatively the experimental results~\cite{Labousse_NJP_2014}. In this article, for the sake of tractability, we take the continuous limit, meaning the above summation yields
\begin{equation}
h\simeq h_0\int\limits_{-\infty}^{t} \frac{dT}{T_F} \; J_0\left(k_F \Vert \vectr(t)-\vectr\left(T\right) \Vert \right)e^{-(t-T)/\tau }.
\label{Bushwave}
\end{equation}
We shall discuss in Sec.~\ref{Lypuexp} the consequences of reaching the continuous limit. This continuous form has been proposed by Mol\'a\v{c}ek \et Bush~\cite{Molacek_JFM_1_2013,Molacek_JFM_2_2013} and yields
\begin{equation}
\ddot{\vectr}= -\tmu \dot{\vectr}-\omega^2\vectr-\tC\left[\bmnabla h\right]_{\vectr}.
\label{EquationofmotionBush}
\end{equation}
$\vectr(t)$ denotes the drop position at time $t$. $\tmu$ stands for an averaged dissipation  friction per unit mass. $-\tC\left[\bmnabla h\right]_{\vectr}$ is the force guiding the drop with a coupling coefficient per unit mass $\tC$. The external force $-m_W \omega^2\vectr$ is applied by means of a magnetic field as described in Sec.~\ref{methods}. The hydrodynamic origin of each coefficient has been investigated in~\cite{Molacek_JFM_2_2013}.\\

We now reformulate the equations of motion into a normal form. The main source of complexity is embedded within the nonlinear integral term as it encodes long-term correlations. A part of the information is encoded into the wave field and the rest of information is stored in the drop position and momentum.  The aim of this paragraph is to clarify what means this interplay of information storage. Which dimensions store wave information? How is information and momentum exchanged between the wave and the drop? And eventually what sets the route toward chaos? \\
\\
Under the integro-differential form (Eq.~\ref{EquationofmotionBush}), the dynamics appears as non-local in time. We first show that this equation can be reformulated as a local normal form  $\dot{\bm{\mathcal{Z}}}= \bm{\mathcal{F}}\left( \bm{\mathcal{Z}}\right)$ where $\bm{\mathcal{Z}}$ is the state vector of the dynamics. This normal form will allow us in section III A 2 to compute the phase space divergence thanks to the partial derivatives of $\bm{\mathcal{F}}$. M. Miskin proposed in~\cite{Perrard_Nature_2014} to decompose the wave terms into a Bessel wave basis. The Graf's decomposition theorem enables one to project each "$J_0$" terms of Eq~\ref{Bushwave} into a Bessel wave basis centred at the origin imposed by the external harmonic potential. The derivation of a local normal form describing the walker dynamics from Eq.~\ref{EquationofmotionBush} is detailed in appendix. We recall here its final form, obtained from a central mode decomposition of the wave field into the radial basis $(r,\theta)$. It yields
\begin{equation}
\left\{
    \begin{array}{ll} 
  \displaystyle     \ddot{x}=-\tmu \dot{x}-\Omega^2 x + \tC k_F h_0\sum\limits_{n\geq 0}C_n\mathcal{T}_{n,x} +S_n\mathcal{V}_{n,x}\\
  \displaystyle     \ddot{y}= -\tmu \dot{y}-\Omega^2 y + \tC k_F h_0\sum\limits_{n\geq 0} C_n\mathcal{V}_{n,y}-S_{n,y}\mathcal{T}_{n,y}\\
       \\
      \dot{C}_n=-C_n/\tau+J_n\left(k_F\sqrt{x^2+y^2} \right)T_n(x,y) \;\;\;\;\ \forall n\geq 0 \\
      \dot{S}_n=-S_n/\tau+J_n\left(k_F\sqrt{x^2+y^2} \right)V_n(x,y)  \;\;\;\;\ \forall n\geq 1
     \end{array}
\right. ,
\label{equationdyninertia}
\end{equation} 
where the functions $T_n$ are the first kind Tchebychev polynomials that can be calculated by recurrence, and $V_n$, $\mathcal{T}_{n,x}$ and $\mathcal{V}_{n,x}$ can be calculated from the $T_n$ and $J_n$ functions (see appendix for details). The mode amplitude $C_n$ and its phase quadrature $S_n$ describes the amplitude and the phase of the mode $J_n$ in the Bessel decomposition of the wave field. They are given by
\begin{equation}
\left\{
    \begin{array}{ll}     
     \displaystyle   C_n=\sum\limits_{k=-\infty}^N J_n\left(k_F r_k\right)e^{-(t_N-t_k)/\tau}\cos\left(n\theta_k \right)\approx \int_{-\infty}^t \frac{dT}{T_F}J_n\left(k_F r(T)\right)e^{-(t-T)/\tau}\cos\left(n\theta(T) \right)\\
        \\
     \displaystyle   S_n=\sum\limits_{k=-\infty}^N J_n\left(k_F r_k\right)e^{-(t_N-t_k)/\tau}\sin\left(n\theta_k \right)\approx \int_{-\infty}^t \frac{dT}{T_F}J_n\left(k_F r(T)\right)e^{-(t-T)/\tau}\sin\left(n\theta(T) \right)         
     \end{array}
\right. .
\end{equation} 
The true interest of the rewriting stands in the clear separation between the evolution of the particle trajectory and its associated wave fields. We can now explicitly write a state vector
\begin{equation}
\bm{\mathcal{Z}}=\begin{bmatrix}
 x \\
\dot{x}\\
y\\
\dot{y}\\
C_0 \\
C_1 \\
S_1 \\
\vdots \\
 \end{bmatrix}
\end{equation}
and remark that the dynamical system of Eq.~\ref{equationdyninertia} formally writes
\begin{equation}
\dot{\bm{\mathcal{Z}}}= \bm{\mathcal{F}}\left( \bm{\mathcal{Z}}\right)
\label{reformulationnormale}
\end{equation} 
The memory information stored in the wave can be stored into a set of modes supporting the wave dimensions of the dynamics. Conceptually, the first four dimensions $ (x,\dot{x},y,\dot{y})$ corresponds to the particle phase space while the other dimensions $(C_0,C_1, S_1\ldots)$ correspond to the wave counterpart. Note that the wave is responsible for supporting an arbitrary large number of additional degrees of freedom in the same spirit as the theoretical investigations of Labousse \etal~\cite{Labousse_These,Labousse_NJP_2014} and of T. Gilet~\cite{Gilet_PRE_2016}. The equation \ref{Forcecarte} indicates a linear relation between $\left[\partial_x h \right]_{\vectr}$ and $\left[\partial_y h \right]_{\vectr}$ and $C_i$ et $S_i$ but a strong non-linear relation with the position coordinates. So the wave dimensions couple linearly to the particle dimensions. In contrast, the particle dimensions couple non-linearly to the wave dimensions. In this Section~\ref{Normalform}, we have translated the path memory integral into a set of modes evolution. Eq.~\ref{reformulationnormale} is still very complex, but this form of equations enables i) to define a phase space and ii) to calculate in the next section some very general properties of the system and its phase space contraction. 
\subsubsection{Phase space contraction \label{Lypuexp} and noise-driven chaos}
In this section we aim at investigating theoretically the phase contraction properties. We consider a small volume of phase space
\begin{equation}
\delta \mathcal{V}_d=d xd \dot{x}d y\delta \dot{y} dC_0 dC_1 dS_1\ldots dC_d dS_d =\prod\limits_{j=1}^{(2d+1)+4} d\mathcal{Z}_i
\label{volumephasespace}
\end{equation}
We first consider an arbitrary large but finite dimension cut-off $d$ so that Eq.~\ref{volumephasespace} and every equations that follow have an unambiguous mathematical meaning. We will consider the limit  $d\rightarrow +\infty$ at the end of the section. The divergence rate of this small phase space volume $\delta \mathcal{V}_d$ towards a given direction is formally related to the Jacobian matrix
\begin{equation}
J_{i,j}=\dfrac{\partial \mathcal{F}_i}{dZ_j}
\end{equation}
Specifically $J_{i,j}$ is the phase space deformation $dZ_j$ towards the direction of phase space $i$. The eigenvalues of $J$ are directly related to the Lyapunov spectrum of the system $\left\lbrace \lambda_i\right\rbrace _{i=1,\ldots, 2d+5}$.  This information depends in general on the given position in the phase space $\bm{\mathcal{Z}}$ and is not an invariant quantity of the dynamics. However we found that it is possible to compute an invariant of the system. Indeed, we recall that the time evolution of the phase space volume is given by~\cite{Lesne,Manneville_2006}   
\begin{equation}
\dfrac{d}{dt}\delta \mathcal{V}_d=\left( \mathrm{div} \mathcal{F} \right)\delta \mathcal{V}_d
\label{eqdiffphasespacevolume}
\end{equation}
The divergence rate $\mathrm{div} \mathcal{F}$ is the trace of the Jacobian matrix $J$ and is base invariant. Its value does not depend of the choice of representation but could depend on position in the phase space $\bm{\mathcal{Z}}$. We find that is not the case and that $\mathrm{div} \mathcal{F}$ is an invariant of the dynamics itself, specifically
\begin{equation}
\mathrm{div} \mathcal{F}=-2\tmu-\dfrac{2d+1}{\tau}.
\label{divFd}
\end{equation} 
Integrating Eq.~\ref{eqdiffphasespacevolume} gives
\begin{equation}
\delta \mathcal{V}_d(t)=\delta \mathcal{V}_d(t=0)e^{\mathrm{div} \mathcal{F} t}
\label{dV}
\end{equation}
First $\mathrm{div}\mathcal{F}$ is constant and always negative, implying that any initial volume of phase space contracts. Note that this does not prevent the existence of diverging directions of phase space, but this means that this divergence will be overcompensated by a contraction in another direction of phase space. Secondly, a contraction of phase space represents a loss of information. This loss has two distinct origins: a mechanical loss of information through the dissipation with the bath and ambient air at rate $-2\tmu$ and a wave loss of information at rate $-(2d+1)/\tau$. Note that the loss of information per mode $1/\tau$ is governed by the sole memory parameter.\\

Finally we calculate the global Lyapunov exponent $\lambda=\sum_i \lambda_i$ of the system. From any arbitrary infinitesimal volume $\delta \mathcal{V}_d$ we can construct an arbitrary infinitesimal distance in the phase space $\delta L$ as 
\begin{equation}
\delta L\propto \delta \mathcal{V}_d^{1/\mathcal{D}}, 
\label{scalinglaw}
\end{equation} 
with the $\mathcal{D} =(2d+1)+4$ the phase space of dimension. Combining Eqs~\ref{dV} and~\ref{scalinglaw} and then taking the limit $d \to +\infty$ we get
\begin{equation}
\delta L(t)=\delta L(t=0)e^{-t/\tau}
\end{equation}
We may now identify the global Lyapunov exponent $\lambda$ of the system as
\begin{equation}
\lambda=-\frac{1}{\tau}
\end{equation}
By taking the limit $d\rightarrow +\infty$ the wave dimensions dominates the scaling law (Eq.~\ref{scalinglaw}) and signifies that the phase contraction is mainly driven by the wave loss of information. Thus the wave dimensions act as an infinite number of additional degrees of freedom. We stress again that $\lambda<0$ does not prevent from chaotic behaviours. Interestingly, for large memory parameter $\tau$, $\lambda$ converges towards 0 with negative values. This convergence to an asymptotic wave neutrality has direct consequences if the dynamics is discrete. Indeed the coarse-graining of the dynamics implies finite amplitude perturbation that can break the stability of an asymptotically neutrally stable attractor. The loss of stability in a phase space of large dimension triggers a route towards chaos. The driven mechanism is the encoding of the finite amplitude perturbation within the infinite wave degrees of freedom. This mechanism will be at play in Sec.~\ref{noisechaos}.  We now turn to the existence of a different fundamental limit triggering a distinct instability mechanism. 

\subsection{General results for wave modes for periodic orbits: a route towards low-dimensional chaotic scenario ~\label{geoconstrain}}
	
The existence of periodic solutions with central symmetry implies mathematical constraints on the dynamics we shall now investigate. These constraints may originate either from the dynamics or the geometry. We first exhibit these different constraints and show their interplay. Finally we show that these constraints lead to a non-linear mechanism promoting unstable situations. 

\subsubsection{Mechanical and geometrical constraints}
We consider an orbit with a central symmetry of period $T_o$. We first derive dynamical constraints. By integrating the equations of motion (Eq.~\ref{EquationofmotionBush}) over one period of time between $t_0$ and $t_0+T_o$, we obtain the first mechanical constrain

\begin{equation}
\int_{0}^{T_0}\left[\bmnabla h\right]_{\vectr} dt =0
\end{equation} 
For symmetry reasons, all the other terms disappear. For periodic orbits, the origin of time $t_0$ is irrelevant and is accordingly set to zero. For periodic orbits, the path integral is a summation of identical but progressively damped sequences. This geometry series simplifies the computation of the path-memory integral to one single orbit through the relation 
\begin{equation}
h= h_0\int\limits_{-\infty}^{t} \frac{dT}{T_F} \; J_0\left(k_F \Vert \vectr(t)-\vectr\left(T\right) \Vert \right)e^{-(t-T)/\tau }=\dfrac{h_0}{1-e^{-T_o/\tau}}\int\limits_{t-T_o}^{t} \frac{dT}{T_F} \; J_0\left(k_F \Vert \vectr(t)-\vectr\left(T\right) \Vert \right)e^{-(t-T)/\tau }
\end{equation}	
So a periodic trajectory must satisfy the mechanical constrain
\begin{equation}
\int_{0}^{T_o}\bmnabla_{\vectr}\int\limits_{t-T_o}^{t} \frac{dT}{T_F} \; J_0\left(k_F \Vert \vectr(t)-\vectr\left(T\right) \Vert \right)e^{-(t-T)/\tau }=0
\label{meca}
\end{equation}
We now exhibit geometrical constraint arising from the periodicity of the paths. After a transient the existence of periodic trajectories imposes that 
\begin{equation}
h(t)-h(t-T_o)=\dfrac{h_0}{T_F}\epsilon(t),
\end{equation}
with $\epsilon(t)\rightarrow 0$ after a typical period of time corresponding to the transient. This means that  
\begin{equation}
\int\limits_{t-T_o}^{t} \frac{dT}{T_F} \; J_0\left(k_F \Vert \vectr(t)-\vectr\left(T\right) \Vert \right)e^{-(t-T)/\tau }=\epsilon(t)
\label{geo}
\end{equation}
We find that the geometrical constrain (Eq.~\ref{geo}) implies that the mechanical constrain (Eq.~\ref{meca}) is always satisfied at long time. Conversely Eq~\ref{meca} requires periodic motion and a radial force. As a contraposition, any non radial external force prevents periodic trajectories. Thus the geometrical constrain (Eq.~\ref{geo}) is more demanding  than the mechanical one (Eq.~\ref{meca}) and will be a preferred starting point for pursuing the calculation. 

\subsubsection{Incompatible wave mode constraints and instability}
Expanding the field in Eq.~\ref{geo} into a Bessel basis leads to 
\begin{equation}
\sum_{n\in \mathbb{Z}}J_n\left(k_F r(t)\right)e^{in\theta (t)}\int\limits_{t-T_o}^{t} dT\; J_n\left(k_F r\left(T\right) \right)e^{-in\theta (T)}e^{-(t-T)/\tau }=\epsilon(t)
\end{equation}
For long memory we get 
\begin{equation}
\sum_{n\in \mathbb{Z}}J_n\left(k_F r(t)\right)e^{in\theta (t)}\int\limits_{t-T_o}^{t} dT\; J_n\left(k_F r\left(T\right) \right)e^{-in\theta (T)}=\epsilon(t)+\mathit{O}\left(\frac{T_0}{\tau}\right)
\end{equation}
We denote $A_n=\int\limits_{t-T_o}^{t} dT\; J_n\left(k_F r\left(T\right) \right)e^{-in\theta (T)}$. We integrate all the terms between $t-T_o$ and $t$ and by symmetry of the integration domains, we get 

\begin{equation}
\frac{1}{2}\sum_{n\in \mathbb{Z}}A_n A_n^*=\tilde{\epsilon}(t),
\label{eqmode}
\end{equation} 
with $\tilde{\epsilon(t)}$ arising from the integration of the right-hand side of Eq.~\ref{eqmode} and converging to $0$ for large time and large memory. Now consider a path invariant under a rotation of angle $2\pi/N$ as observed experimentally. We have $A_n=0$ if $n/N$ is not an integer. So the summation reduces to the modes symmetrically compatible specifically
\begin{equation}
\dfrac{1}{2}\sum_{p\in \mathbb{Z}}\vert A_{p\times N}\vert^2=\tilde{\epsilon}(t)
\label{eqmodefinal}
\end{equation} 
As thoroughly investigated for circular paths~\cite{Oza_JFM_2_2013,Labousse_PRE_2016}, Eq.~\ref{eqmodefinal} means that the mode $0$ extinguishes, $\vert A_{0}\vert \simeq 0$ at high memory. For higher order symmetries, say a lemniscate (N=2), Eq.~\ref{eqmodefinal} implies that $\vert A_{0}\vert \simeq 0$ and $\left\lbrace \vert A_{\pm 2k}\vert \simeq 0\right\rbrace_k$ simultaneously. The size quantisation of the higher order solutions of the dynamics originates in this simultaneous mode extinction. The rather limited number of constraints involved supports a low-dimensional mechanisms as described in Sec.~\ref{chaosdet}. However, at high memory, as $\tilde{\epsilon}\rightarrow 0$, these constraints are difficult to be simultaneously satisfied at all time, which leads to low-dimensional instabilities. We now turn back to the experimental characterisation of these different scenarios.

	\section{Experimental characterisation of the different chaotic scenario~\label{expchar}}

In this section, we seek for deterministic behaviour within the  intermittent trajectories obtained experimentally in a harmonic well. A unified picture of the transition to chaos in wave-memory dynamics in confined situation has not been achieved. Instead, two regimes qualitatively different have been observed. We will present the two archetypes of intermittency types that have been observed for various spatial extension of the trajectories and different symmetry of the modes involved. It concerns the intermittency within the single lemniscate mode $(n,m)=(2,0)$, the intermittency between circular orbits $n=1$ and the lemniscate mode $(n,m)=(2,0)$, and eventually the intermittency between circular orbits at $n=2$.\\

To look for order in disorder, one canonical method is the use of first return map. The main idea is to reduce the description of a dynamics of a continuous variable in time $X(t)$ to a dynamics of a variable discrete in time $X_k$~\cite{Floquet_1883}. We then look for an expression of $X_{k+1}$ as a function of $X_k$. For a quasi-periodic system in which the main temporal evolution is an oscillation at period $T$ with additional slow variation in time one can naturally chose $X_k = X(k T)$. From an experimental point of view, the period $T$ may be ill-defined. In practice, the choice of the local extremum of $X(t)$ for the $X_k$ is more suitable to identify deterministic behaviour~\cite{Abarbanel}. The time interval $T$ between $X_k$ and $X_{k+1}$ can therefore vary with time as it is self-determined by the duration between two successive extrema. We then seek to describe the time serie $X_k$ by the mean of a function $G$ such that :
	\begin{equation}
	X_{k+1} = G(X_k)
	\end{equation}
	If $G$ exists it is called the application of first return. This iterative point of view has been widely used in the 70's and 80's to describe the temporal evolution of highly disordered temporal signals~\cite{Eckmann_1981}. The first experimental use of the first return map has been used in particular on a model of thermal convection by Lorenz who identified a chaotic behaviour in a system of low dimensionality~\cite{Lorenz_1963}.
\subsection{Experimental characterisation low-dimensional deterministic chaos~\label{chaosdet}}
\subsubsection{Lemniscate $n$ = 2 :  Chaos of pure lemniscate}	
\begin{figure}
\centering
\includegraphics[width=0.95\columnwidth]{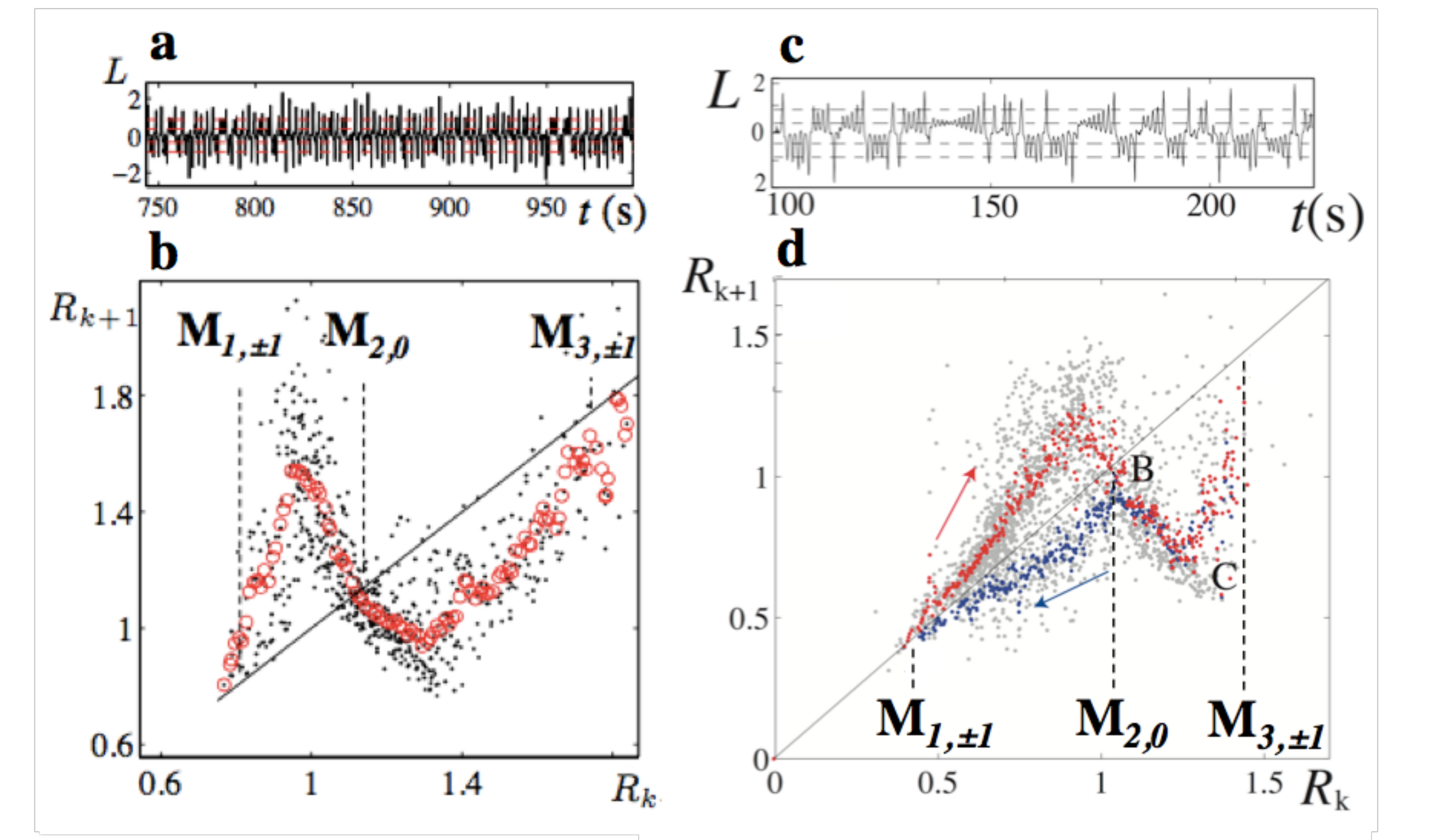}
\caption{{\bf First return map} a) Temporal record of dimensionless angular momentum $L$ obtained for $\Lambda = 0.57$ and $V = 8.6$~mm/s for $M=100$. b) Associated first return map (single valued case). The local maximum $R_{k+1}$ is represented as a function of the previous maximum $R_k$. Ensemble average on $R_k$ values for bins of width $\delta R = 0.005$ (\textcolor{red}{$\circ$}). Fix point M$_{2,0}$ of the first return map defined by the intersection with $R_{k+1}=R_k$ (solid line). c) Temporal record of $L$ for $\Lambda = 0.49$ and $V = 9.7$~mm/s for $M=63$ using also the local maximum $R_k$. d) Associated first return map (bi-valued case) for $R_k$. Ensemble average on $R_k$ values for bins of width $\delta R = 0.005$ (\textcolor{red}{$\circ$}) for $R_k>R_{k-1}$ (\textcolor{red}{$\cdot$}) and $R_k < R_{k-1}$ (\textcolor{blue}{$\cdot$}). The two branches are well separated between A and B, and mixed between B and C. Fix point M$_{1,\pm 1}$ and M$_{2,0}$ of the first return map define by the intersection with $R_{k+1}=R_k$ (solid line). Fig~3c \& 3c adapted from~\cite{Perrard_PRL_2014} with courtesy of Perrard~\cite{et al.}}
\label{fig3}
\end{figure}
For sake of simplicity we first analyse the low-dimension chaos type. We build a first return map from the temporal record of the distance $r$ to the center. Figure~\ref{fig3} shows two realisations computed from a lemniscate (figures~\ref{fig3}a and b) and from an intermittency between circular orbits $n=1$ and lemniscate ($n=2$) mode (figures~\ref{fig3}c and d). The temporal record of the angular momentum has been added to evidence the transition between modes of different symmetry. We first focus on the case of a single mode shown in figures~\ref{fig3}a and b. It was obtained with a confinement $\Lambda = 0.57$ so in the vicinity of the lemniscate stability range. The first return map has been computed from the successive maximum $R_k$ of the distance $r$ to the center. Each black dot corresponds to a measure while the red circles were obtained from an average of all dots in the range $R$ and $R+\delta R$ with $\delta R = 0.005 \lambda_F$. As we can see on figure~\ref{fig3}b, the experimental realisation of $G$ is single valued : one can indeed describes the dynamics with a deterministic function $G$. The fixed points of the map $G_{\mathrm{exp}}$ namely M$_{1,\pm 1}$, M$_{2,0}$ and M$_{3,\pm 1}$ correspond to attractors of the dynamics. In the current case, none of them are stable. Most of the dynamics occurs in the vicinity of M$_{2,0}$, but excursions near M$_{1,\pm 1}$ and M$_{3,\pm 1}$ have also been observed which correspond in real space of loops of variable extensions and orientations. As a consequence, the apparent signal is disordered even if there exists an underlying deterministic evolution.
		
\subsubsection{Orbits $n$ = 1 : deterministic chaos, type III.}

	The intermittency near the stability range of $n=1$ orbits can also be studied using the first return map although it exhibits qualitative difference. Figure~\ref{fig3}d extracted from~\cite{Perrard_PRL_2014} shows the first return map built from the successive maximum $R_k$ of the distance to the center for $\Lambda = 0.49$ and $M=63$. Each grey dot corresponds to a measure of one period of motion. In a similar manner, it shows the existence of three fixed points M$_{1,\pm 1}$, M$_{2,0}$ and M$_{3,\pm 1}$ along the line $R_{k+1} = R_k$. The associated trajectory in the vicinity of these points corresponds respectively to the modes $(1,\pm 1)$, $(2,0)$ and $(3,\pm 1)$. The striking aspect of this first return map is the presence of two branches between M$_{1,\pm 1}$ and M$_{2,0}$ located on each side of the line $R_{k+1} = R_k$. These two branches can however be distinguished by adding the previous maximum radius $R_{k-1}$ as a variable. The conditional average on $R_{k}>R_{k_1}$ gives the red dots curve, while the conditional average on $R_k < R_{k-1}$ gives the blue dots curve. The red branch of our first return map is reminiscent of the Lorenz attractor first return map~\cite{Lorenz_1963} or the R\"ossler attractor~\cite{Rossler_1978}. The high slope of $G$ near its maximum makes this dynamics highly sensitive to the initial conditions. If one can predict the temporal evolution starting from any initial condition this predictability is in practice lost after a finite time.
		
	The presence of two branches in this application breaks the deterministic evolution of a single variable $R_k$. This predictability can be recovered by adding the new variable $R_{k-1}$. However, this choice introduces temporal correlation in the dynamics and thus does not correspond to a first return map. In order to build a two dimensional first return map without invoking temporal correlations, one has to look for a new relevant variable $Y_k$ describing the system state at instant $k$. Neither the angular position of the speed nor the drop position turn out to be relevant. In order to keep a description local in time of the instant $k+1$ as a function of instant $k$, one has to add a variable which is not associated $\bf v$ or $\bf r$. One way to rationalise the evolution is to choose for $Y$ a variable associated to the wave field surrounding the drop. Following the Graf's theorem decomposition introduced in section III, a natural choice could be the amplitude $A_0$ of the centred mode $J_0$.	
	
	Eventually, there exists two routes to rationalise this multivalued first return map. The first one includes variables associated to the wave field. Doing so, we introduce a field variable for the motion of a localised object. The dynamics is then no more equivalent to the motion of a localised particle in 2 dimensions of space. The second rationalising path adds a variable of the previous time $R_{k-1}$. It is still described solely by the motion of a localised particle in space, but the temporal locality is lost. In that sense, the multivalued first return map of figure~\ref{fig3} is a signature of the \textit{wave memory}, that makes this dynamics either non local in time or in space.

\subsection{Experimental characterisation of effective noise-driven chaos~\label{noisechaos}}
\begin{figure}
\centering
\includegraphics[width=0.7\columnwidth]{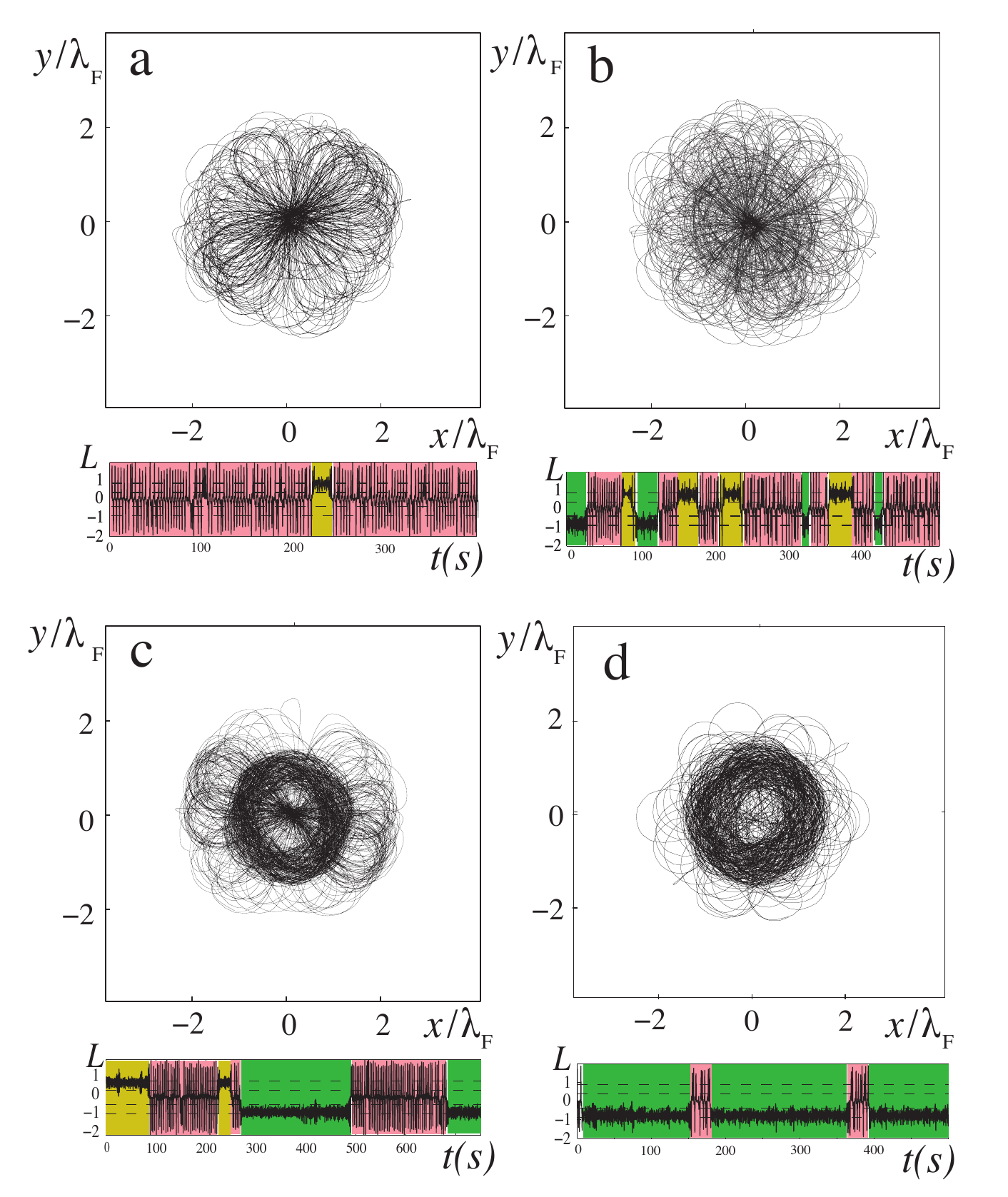}
\caption{{\bf Intermittency of circular orbits $n=2$} Trajectory and angular momentum $L$ for increasing values of $\Lambda$, $M = 50$ and $V = 11.5 \pm $0.3~mm/s. a) $\Lambda = 0.73$. b) $\Lambda = 0.75$. c) $\Lambda=0.76$. d) $\Lambda =0.77$.}
\label{fig4}
\end{figure}
The next parameter range that exhibits intermittency behaviour is located between the stability range of the lemniscate mode $(n,m)=(2,0)$ and the circular modes $(n,m) = (2,\pm 2)$ for $0.65 < \Lambda < 0.8 $. Figure~\ref{fig4} shows four examples of trajectories obtained within this range for a memory parameter $M = 50$ and increasing values of $\Lambda$. The temporal record of the instantaneous angular momentum $L_z$ also shows evidence of a dynamics switching between values that correspond to eigenmodes $(n,m)$ of the dynamics (dashed lines). The time spent in any of these modes can last hundreds of periods mediated by fast erratic transitions. Contrary to the previous cases around $n=1$ and $(n,m) = (2,0)$, no slow growth of a wobbling amplitude have been observed. The first return map have been computed for the distance to the center $R$, the orientation $\theta$, the angular moment $L_z$ and the amplitudes of the first modes of the wave field Bessel decomposition. None of them have revealed a deterministic function $G$. These maps only evidence an accumulation of points around unstable fixed points, without deterministic link between successive events. We did not succeed in revealing a global underlying deterministic behaviour. \\
	
\begin{figure}
\centering
\includegraphics[width=0.6\columnwidth]{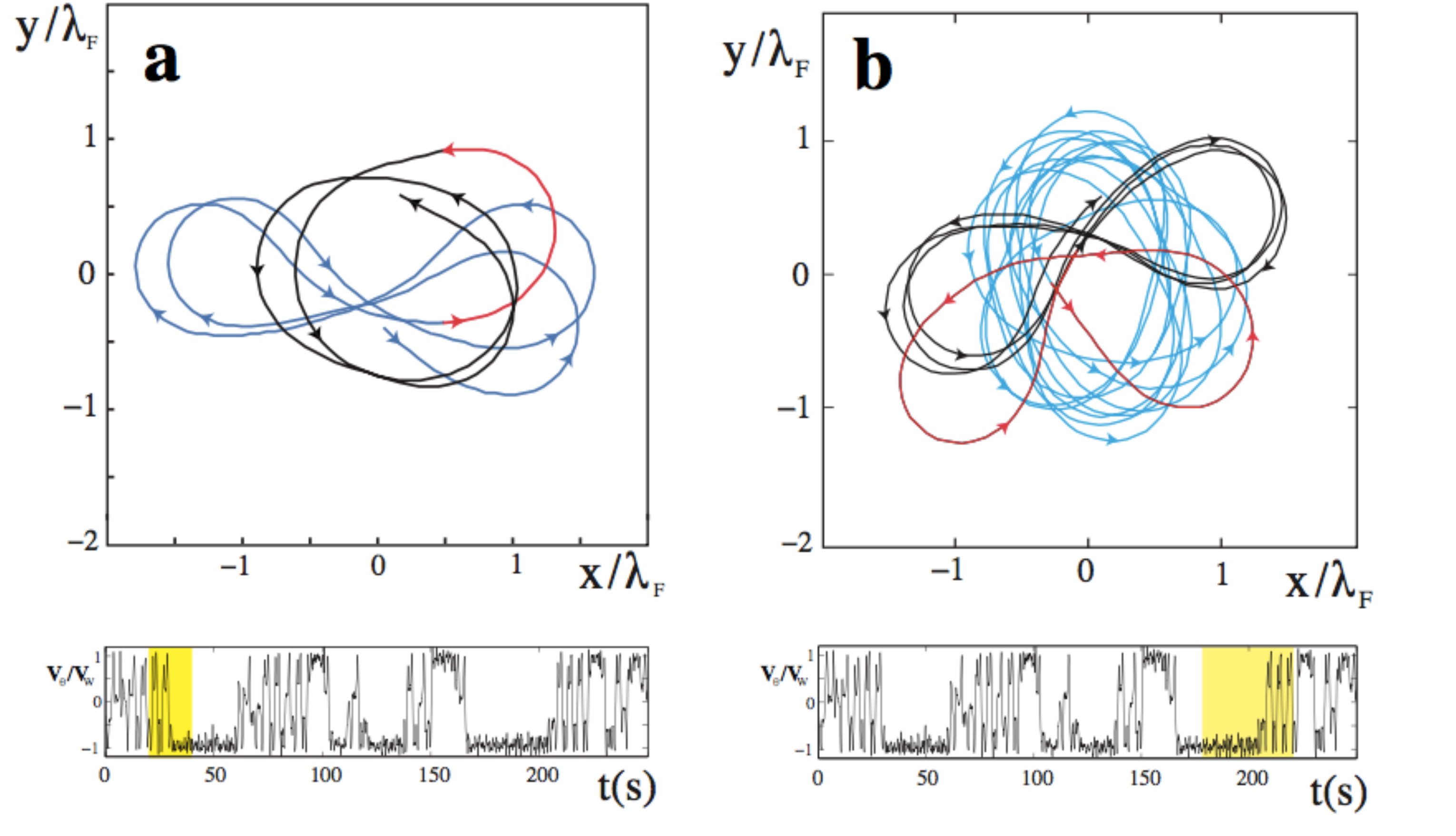}
\caption{{\bf Archetype of transition between modes $n=2$.} Trajectory and angular momentum near a spontaneous transition at the onset of intermittency ($M = 25$, $\Lambda = 0.62$ and $V = 7.8$~mm/s). Both transitions are extracted from the same recording.Trajectory before the transition (blue) during the transition (red) and after the transition (black). a) Transition $(n,m)=(2,0) \rightarrow (2,-2)$. b) Transition $(n,m)=(2,-2) \rightarrow (2,0)$.}
\label{fig5}
\end{figure}

	The failure of the deterministic description can originate from an increase of the number of relevant degrees of freedom necessary to describe the dynamics. Indeed without any apriori symmetry consideration, the number of excited mode increases (as a square roots) with the orbit size. This increase of relevant dimensions is known to be an obstacle for the search of determinism in experimental realisations of a chaotic system~\cite{Manneville_2006}. However, even if the global deterministic description is apparently lost, the transition between modes still exhibit reproducible patterns. The two archetype transitions for $(2,0) \rightarrow (2,2)$ and $(2,2) \rightarrow (2,0)$ are shown in figure~\ref{fig5}. From a lemniscate mode, an excursion of large amplitude generates one loop of larger radius of curvature that stay away from the center (red line of figure ~\ref{fig5}a). Once one large loop has been achieved, the drop keeps moving along a stadium shape characteristic of $(n,m)=(2,\pm 2)$ modes. The transition back from stadium to lemniscate involves a transient two successive loops of small radius of curvature passing by the centre of the harmonic well~(figure~\ref{fig5}b). Once a loop of opposite angular momentum has been drawn, the drop keeps moving in a lemniscate mode until it will destabilise again. Due to the iterative characteristic of the walker's dynamics, the observation of these trajectories presents another interest. As can be observed from Eqs.~\ref{Fortwave}\&\ref{EquationofmotionBush}, the drop trajectory encodes the entire information necessary to reconstruct the dynamics. The two transitions of figure~\ref{fig5} are therefore remarkable examples of representation of a bistable dynamics in real space. \\
		
	A dynamics composed of an erratic evolution near the attractors with fast deterministic behaviour between attractors can be reminiscent of low-dimensional systems coupled to an external noise. It is in particular the case for magnetic field earth reversal in which the magnetic field is submitted to a multiplicative noise throughout the coupling with the turbulent velocity field~\cite{Berhanu_EPL_2007,Fauve_2013}. Depending on the nature of stable and unstable fixed points, low-dimensional systems coupled to an external noise can also be associated to ON-OFF intermittency~\cite{Platt_1993,Heagy_1994}. The low number of relevant dimensions, the reproducibility of transient regimes and the existence of a multiplicative noise are common features of these system. We will see that a multiplicative noise is also found in walker's dynamics.
	
In the walker dynamics, there is a source of multiplicative noise which amplitude may grow with the memory parameter. The temporal variability of the impact time and impact positions generate a fluctuation of the wave amplitude $h_0(1+\eta_k)$ where $\eta_k \ll 1$ is the noise amplitude. The total wave force exerted on the drop writes :
\begin{equation}
\mathbf{F}_{\mathrm{waves}}  \propto - {\bs \nabla}_{\bs r_N} \sum_{k=-\infty}^{N} h_0 (1 + \eta_k) J_0 (k_F || {\bs r}_N - {\bs r}_k ||) e^{-(t_N-t_k)/\tau}
\label{eq_noise1}
\end{equation}
	If we consider \textit{e.g.} the $\eta_k$ as independent random variables with $\left\langle \eta_k \right\rangle =0$ and $\left\langle\eta_k^2\right\rangle = \eta^2$, the central limit theorem applies and the total fluctuating part of the wave force $\delta F_{waves}$ writes :
\begin{equation}
\mathbf{\delta F}_{\mathrm{waves}} \approx \eta \sqrt{M} \bs F_{waves}
\label{eq_noise2}
\end{equation}
	Even if the noises $\eta_k$ on each individual source is of negligible amplitude, the temporal recording in the wave field accumulate noise effects. Numerical simulations performed with noise of increasing amplitude have previously revealed its key influence on the dynamics~\cite{Labousse_PRE_2_2016}. The temporal recording of noise effects in the wave field and its influence on the dynamics is still an open question that would need further investigations. The accumulation of noises and its multiplicative nature, as well as the temporal record of angular momentum shows strong evidences of an analogy with multi-stability cases driven by multiplicative noises. The existence of two modes of same symmetry $(n,m)= (2,\pm 2)$ whose transitions are mediated by a more unstable mode of different symmetry is reminiscent of the magnetic earth reversal~\cite{Petrelis_2009}. \\

	When one observes more extended trajectories for increasing values of $\Lambda$, the complexity of the trajectories and the number of stable modes involved keeps increasing. As a result, no deterministic behaviour has been identified. A route to chaos as described numerically by Tambasco~\textit{et al.}~\cite{Tambasco_2016} is more likely to happen. In this particular case, the absence of frequency doubling cascade in the presence of a harmonic well is an argument for the Ruelle-Takens-Newhouse scenario~\cite{Ruelle_1971,Newhouse_1978}. However, the existence of a multiplicative noise of increasing amplitude with memory may limit, in practice, the observation range of deterministic chaos in the experiments performed with walkers. The influence of noise effects on the dynamics have been directly pointed out~\cite{Labousse_PRE_2_2016}. Noise amplification will always eventually arise due to the amplification mechanism we present here, but the memory limit at which it dominates the dynamics depends on setups and initial noise level to be amplified. A direct study of noises encoding and its influence on the dynamics is still to explored.  \\
	
\begin{figure}
\centering
\includegraphics[width=0.6\columnwidth]{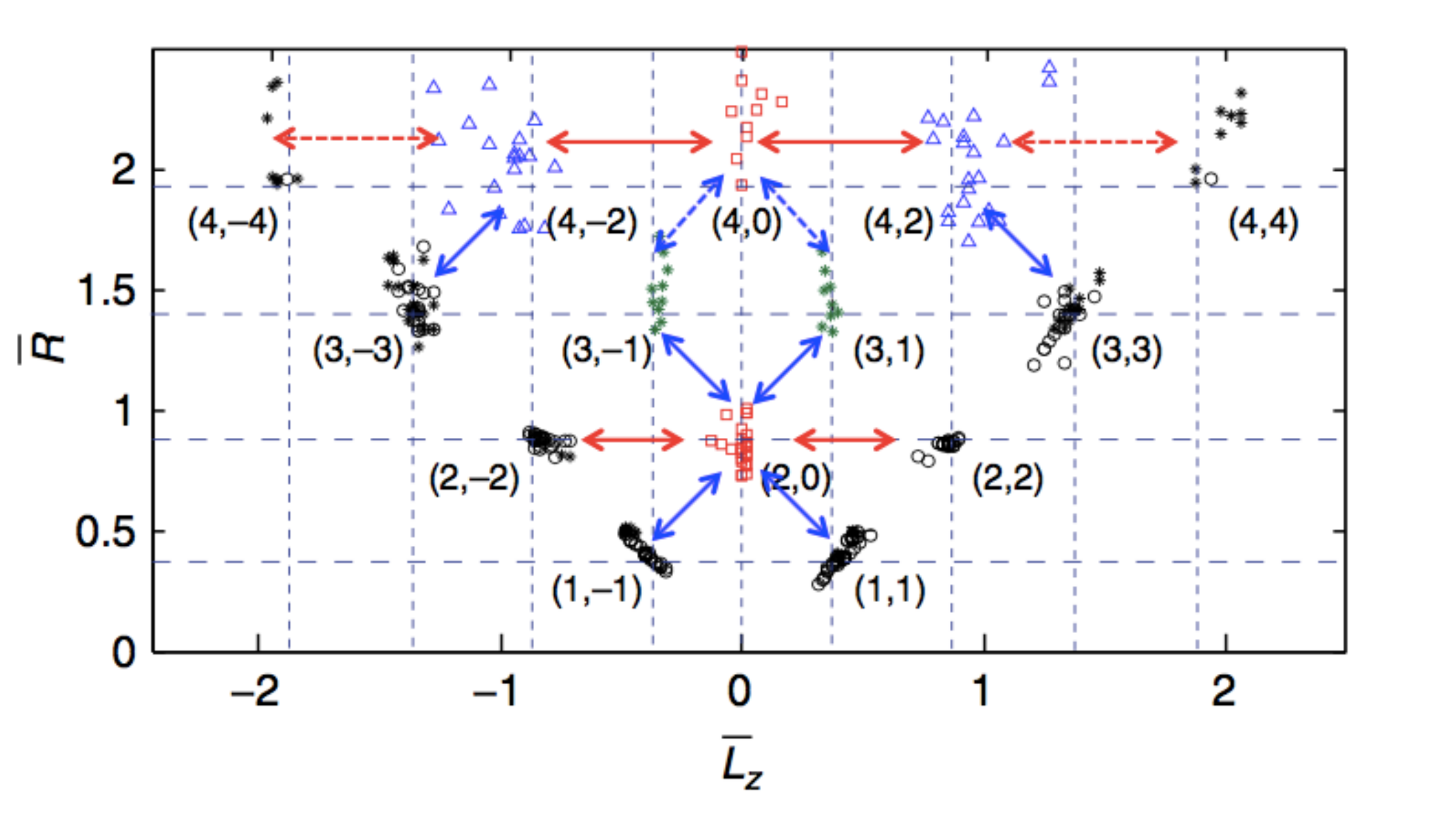}
\caption{{\bf Transitions between levels : an overview.} Diagram in the representation $(\bar R, \bar L)$ of stable trajectories in the shape of circle (black circles), lemniscates (red squares) and trifolium (blue triangles) for $M > 50$. Unstable trajectories that mediate the transition between modes have been added. Arrows represent the documented spontaneous transitions between modes observed for $M>50$.}
\label{fig6}
\end{figure}
As a summary to our experimental investigation, we show in figure~\ref{fig6} the diagram of all observed transitions between modes. The mean angular momentum $\bar L$ is represented as a function of the mean spatial extension $\bar R$ for all the stable trajectories observed experimentally in a harmonic well. Reminiscent long loops observed in chaotic regimes have also been added on this diagram as they mediate the transition between the other modes $(n,m)$. The possible transitions are represented with left and right arrows. Solid lines are documented archetypes while dashed lined would need further experiments to be characterised. It is remarkable to note that all transitions are observed to be local in this representation. They involved either a shift of symmetry $m \rightarrow m \pm 2$ (red lines) or a shift of symmetry and spatial extension $(m,n) \rightarrow (n\pm1,m\pm 1)$ (blue lines). 
	
\section{Conclusion}\label{sec:conclusion}

	In this article we have presented the archetypes of transitions to chaos in walker dynamics under central force confinement, and we have discussed the different scenario leading to disordered trajectories. Section~\ref{expchaos} has presented how to obtain a central force field confinement for a bouncing drop, and present the first sign of instabilities we observed experimentally. Then in Sec.~\ref{theory} we have presented a theoretical investigation of walker dynamics stability in a central force field. We emphasize in particular two possible nonlinear mechanisms originating the transition to disordered trajectories. The first one is based on an analysis of the phase space contraction in the limit of large memory. The global Lyapunov exponent $\lambda$ converges toward 0 from negative values at high memory. Moreover this exponent is dominated by the wave field that contains most of the relevant dimensions. In the high memory limit, any finite size perturbations lead to instability and may trigger chaotic behaviours. In that scenario the discreteness of the dynamics that is encoded within the waves degrees of freedom acts as a noise source and may play a crucial role. The second scenario is based on an analysis of the constraints imposed on the wave field by periodic trajectories. In the high memory limit, we show that the wave mode amplitude must all vanish for ensuring periodic motion. For orbits of small extension it implies only a limited number of relevant constraint. In that regime, the walker dynamics may be described by a small number of degrees of freedom and thus a low dimensional chaotic behaviour is expected. \\
	 
	 In section~\ref{expchar}, we present the experimental characterisation of disordered trajectories. Depending on spatial extension trajectories, two scenario have been identified. The first scenario is a low dimensional chaotic behaviour that can be rationalised using first return map. This map can be defined either from the drop position only (intermittency of lemniscate mode) or from a combination of drop position variables and degrees of freedom associated to the wave field (intermittency between $n=1$ modes and lemniscate). It is in accordance with a low dimensional chaos intuited in section III. The second scenario is a loss of full determinism throughout noise amplification. The absence of experimental deterministic first return map for higher extension modes ($\Lambda > 0.7$) is in favour of this second scenario. In that scenario, the loss of stability is triggered by noise fluctuations. Once the motion has left a stable region of the phase space, a fast and deterministic transition occurs which brings the dynamics in the vicinity of another attractor of the phase space. The noise amplification mechanism detailed in Eqs.~\ref{eq_noise1} and~\ref{eq_noise2} is a key ingredient for understanding walker dynamics in the high memory limit. Finally, we present a diagram of all the observed transitions which shows that the symmetry change during a transition are local, \textit{i.e.} only local shift of symmetry $m \rightarrow m \pm 2$ or local shift of symmetry and spatial extension $(m,n) \rightarrow (n\pm1,m\pm 1)$ have been observed. The emergence of selection rules for confined walkers have been already evidenced in Coriolis force experiments~\cite{Fort_PNAS,Oza_JFM_2_2013} or in central force configuration~\cite{Perrard_Nature_2014,Durey_2017,Kurianski_2017}. We highlight here few mechanisms of transitions from stable orbits toward disordered states from an experimental and a theoretical point of view. It is remarkable that a macroscopic wave particle exhibits also selection rules for transitions between eigenstates. 
	 	
\section{Acknowledgements}\label{sec:acknowledgements}
	The authors thank M. Berhanu, Y. Couder, E. Fort and F. Petrelis for fruitful discussions. S.P. and M.L. acknowledge LASIPS program, the financial support of the French Agence Nationale de la Recherche, through the project ANR Freeflow€™, LABEX WIFI (Laboratory of Excellence ANR-10-LABX-24), within the French Program €˜Investments for the Future€™ under reference ANR-10-IDEX-0001-02 PSL.

\appendix*
\section{Derivation of the ODEs for pilot-wave dynamics}

	We detail here the derivation of Eq.~\ref{equationdyninertia} from the integro-differential form given by Eq.~\ref{EquationofmotionBush}. The Graf's decomposition theorem enables one to project each "$J_0$" terms of (Eq~\ref{Bushwave}) into a Bessel wave basis centred at the origin imposed by the external harmonic potential. This central mode decomposition into the radial basis $(r,\theta)$ yields
\begin{small}
\begin{equation}
\int\limits_{-\infty}^{t}\frac{dT}{T_F}\; J_0\left(k_F \Vert \vectr(t)-\vectr\left(T\right) \Vert \right)e^{-(t-T)/\tau }=\sum\limits_{n=-\infty}^{+\infty}J_n\left(k_F r(t) \right) \int\limits_{-\infty}^{t}\frac{dT}{T_F}\; J_n\left(k_F r\left(T\right) \right)e^{-(t-T)/\tau }e^{in(\theta(t)-\theta(T))},
\end{equation}
\end{small}
and the surface field rewrite
\begin{equation}
h=h_0\sum\limits_{n=0}^{+\infty}\left(2-\delta_{n,0}\right) J_n\left(k_F r \right) \left(C_n \cos\left(n\theta \right)+S_n \sin\left(n\theta \right) \right), 
\end{equation}
with $\delta_{n,0}$ the Kronecker symbol. The mode $C_n$ and its phase quadrature $S_n$ are given by
\begin{equation}
\left\{
    \begin{array}{ll}     
     \displaystyle   C_n=\sum\limits_{k=-\infty}^N J_n\left(k_F r_k\right)e^{-(t_N-t_k)/\tau}\cos\left(n\theta_k \right)\approx \int_{-\infty}^t \frac{dT}{T_F}J_n\left(k_F r(T)\right)e^{-(t-T)/\tau}\cos\left(n\theta(T) \right)\\
        \\
     \displaystyle   S_n=\sum\limits_{k=-\infty}^N J_n\left(k_F r_k\right)e^{-(t_N-t_k)/\tau}\sin\left(n\theta_k \right)\approx \int_{-\infty}^t \frac{dT}{T_F}J_n\left(k_F r(T)\right)e^{-(t-T)/\tau}\sin\left(n\theta(T) \right)         
     \end{array}
\right.
\end{equation} 
The true interest of the rewriting stands in the link between $C_n$ (or $S_n$) and its time derivative
\begin{equation}
\left\{
    \begin{array}{ll}     
        \dot{C}_n=-C_n/\tau+J_n\left(k_F r \right)\cos n\theta\\
        \\
        \dot{S}_n=-S_n/\tau+J_n\left(k_F r \right)\sin n\theta         
     \end{array}
\right.
\end{equation}
 This decomposition in a central wave basis is the first but crucial step toward writing the dynamics in a form local in time. This set of equation means that the modes $C_n$ and $S_n$ support all the waves information. The last step of reformulation consists in expressing the equation of motions into a Cartesian basis 
 
\begin{equation}
\left\{
    \begin{array}{ll} 
       \ddot{x}=-\tmu \dot{x}-\Omega^2 x -\tC\left[\partial_x h \right]_{\vectr}\\
       \ddot{y}=-\tmu \dot{y}-\Omega^2 y -\tC\left[\partial_y h \right]_{\vectr}
     \end{array}
\right.
\end{equation}
The $x$ and $y$ component of the wave force $\left[\partial_x h \right]_{\vectr}$ and $\left[\partial_y h \right]_{\vectr}$ are given by
\begin{equation}
\left\{
    \begin{array}{ll} 
       \left[\partial_x h \right]_{\vectr}= \left(\cos\theta \nabla_r h-\sin\theta \nabla_{\theta} h\right)=-k_F h_0 \left[J_1 T_1C_0+\sum\limits_{n\geq1}C_n\mathcal{T}_{n,x} +S_n\mathcal{V}_{n,x}\right]\\
       \left[\partial_y h \right]_{\vectr}= \left[\sin\theta \nabla_r h+\cos\theta \nabla_{\theta} h\right]=-k_F h_0 \left[J_1 V_1C_0+\sum\limits_{n\geq1} C_n\mathcal{V}_{n,y}-S_n\mathcal{T}_{n,y}\right]
     \end{array}
\right.
\label{Forcecarte}
\end{equation}
with 
\begin{equation}
\left\{
    \begin{array}{ll} 
  \mathcal{T}_{n,x}=J_{n+1}T_{n+1}-J_{n-1}T_{n-1}\\
  \mathcal{T}_{n,y}=J_{n+1}T_{n+1}+J_{n-1}T_{n-1}\\
  \mathcal{V}_{n,x}=J_{n+1}V_{n+1}-J_{n-1}V_{n-1}\\
  \mathcal{V}_{n,y}=J_{n+1}V_{n+1}+J_{n-1}V_{n-1}\\
     \end{array}
\right.
\label{relationTV}
 \end{equation}
 and
\begin{equation}
\left\{
    \begin{array}{ll} 
        \cos n\theta =T_n(cos\theta)=T_n(x,y)\\
        \sin n\theta=V_n(cos\theta)=V_n(x,y)
     \end{array}
\right.
\label{polairetocart}
\end{equation}
The functions $T_n$ are the first kind Tchebychev polynomials that can be calculated by recurrence. Finally the integro-differential equation can be rewritten into the desired local normal form 
\begin{equation}
\left\{
    \begin{array}{ll} 
  \displaystyle     \ddot{x}=-\tmu \dot{x}-\Omega^2 x + \tC k_F h_0\sum\limits_{n\geq 0}C_n\mathcal{T}_{n,x} +S_n\mathcal{V}_{n,x}\\
  \displaystyle     \ddot{y}= -\tmu \dot{y}-\Omega^2 y + \tC k_F h_0\sum\limits_{n\geq 0} C_n\mathcal{V}_{n,y}-S_{n,y}\mathcal{T}_{n,y}\\
       \\
      \dot{C}_n=-C_n/\tau+J_n\left(k_F\sqrt{x^2+y^2} \right)T_n(x,y) \;\;\;\;\ \forall n\geq 0 \\
      \dot{S}_n=-S_n/\tau+J_n\left(k_F\sqrt{x^2+y^2} \right)V_n(x,y)  \;\;\;\;\ \forall n\geq 1
     \end{array}
\right.
\end{equation}

\bibliography{bibliogouttes}

\end{document}